%% file: main.tex
\documentclass[sigconf]{acmart}
\usepackage{xspace}
\usepackage[table]{xcolor}   % 必须带 [table] 选项
\usepackage{booktabs}
\usepackage{multirow}
\usepackage{CJKutf8}      % pdfLaTeX 中文
\usepackage{tabularx}
\usepackage{array}
\usepackage{makecell}   % 导言区
\usepackage{ragged2e}

\definecolor{headblue}{RGB}{40,70,130}
\definecolor{rowgray}{RGB}{240,242,247}
\newcolumntype{C}{>{\centering\arraybackslash}m{2.6cm}}
\newcolumntype{L}{>{\RaggedRight\arraybackslash}X}   % X 默认就是 m 类型(垂直居中)

\newcommand{\zh}[1]{\begin{CJK}{UTF8}{gbsn}#1\end{CJK}}

\newcommand{\eg}{\textit{e.g.}\xspace}

\AtBeginDocument{%
  }

\setcopyright{acmlicensed}
\copyrightyear{2018}
\acmYear{2018}
\acmDOI{XXXXXXX.XXXXXXX}
\acmConference[Conference acronym 'XX]{Make sure to enter the correct
  conference title from your rights confirmation email}{June 03--05,
  2018}{Woodstock, NY}
\acmISBN{978-1-4503-XXXX-X/2018/06}

\begin{document}

% \title{TSGR: Value-aware Generative Retrieval in Taobao Search}
\title{TSGR: Taobao Search Generative Retrieval}

% \affiliation{%

\author{Tianyu Zhan}
\authornote{Equal contribution. Work is done during Tianyu Zhan's internship at Alibaba Group.}
\affiliation{%
  \institution{Zhejiang University}
  \city{Hangzhou}
  \country{China}}
\email{yuzt@zju.edu.cn}

\author{Gui Ling}
\authornotemark[1]
\affiliation{%
  \institution{Taobao \& Tmall Group of Alibaba}
  \city{Hangzhou}
  \country{China}}
\email{linggui.lg@taobao.com}

\author{Tong Xiong}
\authornotemark[1]
\affiliation{%
  \institution{Taobao \& Tmall Group of Alibaba}
  \city{Hangzhou}
  \country{China}}
\email{xiongtong.xt@taobao.com}

\author{Kunhai Lin}
\authornotemark[1]
\affiliation{%
  \institution{Taobao \& Tmall Group of Alibaba}
  \city{Hangzhou}
  \country{China}}
\email{linkunhai.lkh@taobao.com}

\author{Yang Wang}
\affiliation{%
  \institution{Taobao \& Tmall Group of Alibaba}
  \city{Hangzhou}
  \country{China}}
\email{wy416570@taobao.com}

\author{Kaixuan Zhang}
\affiliation{%
  \institution{Taobao \& Tmall Group of Alibaba}
  \city{Hangzhou}
  \country{China}}
\email{zhangkaixuan.zkx@taobao.com}

\author{Zhihong Chen}
\affiliation{%
  \institution{Taobao \& Tmall Group of Alibaba}
  \city{Hangzhou}
  \country{China}}
\email{jhon.czh@taobao.com}

\author{Yuliang Yan}
\authornote{Corresponding author.}
\affiliation{%
  \institution{Taobao \& Tmall Group of Alibaba}
  \city{Hangzhou}
  \country{China}}
\email{yuliang.yyl@taobao.com}

\author{Dan Ou}
\affiliation{%
  \institution{Taobao \& Tmall Group of Alibaba}
  \city{Hangzhou}
  \country{China}}
\email{oudan.od@taobao.com}

\author{Shengyu Zhang}
\authornotemark[2]
\affiliation{%
  \institution{Zhejiang University}
  \city{Hangzhou}
  \country{China}}
\email{sy_zhang@zju.edu.cn}

\author{Haihong Tang}
\affiliation{%
  \institution{Taobao \& Tmall Group of Alibaba}
  \city{Hangzhou}
  \country{China}}
\email{piaoxue@taobao.com}

\author{Bo Zheng}
\affiliation{%
  \institution{Taobao \& Tmall Group of Alibaba}
  \city{Hangzhou}
  \country{China}}
\email{bozheng@taobao.com}

\renewcommand{\shortauthors}{Trovato et al.}

\begin{abstract}
\input{Sections/0_Abstract}
\end{abstract}





\maketitle

\input{Sections/1_Introduction}
\input{Sections/2_RelatedWorks}
\input{Sections/3_Methodology}

\input{Sections/4_Experiment}
\input{Sections/5_Conclusion}
\bibliographystyle{ACM-Reference-Format}
\bibliography{main}

\clearpage
\appendix
\input{Sections/6_Appendix}

\end{document}

%% file: Sections/0_Abstract.tex
Generative retrieval (GR) has demonstrated strong promise for industrial e-commerce search by training a single autoregressive model to directly generate the Semantic IDs (SIDs) of target items.
However, existing GR systems are primarily optimized for semantic matching and remain insensitive to item business value: SID construction is value-unaware, and candidates are ranked without access to item side-info.
Consequently, high-value items are often missed or deprioritized at the retrieval stage, limiting downstream business impact.
This limitation is particularly critical in industrial settings such as Taobao Search, where business objectives are central to system design.
To address this, we propose \textbf{T}aobao \textbf{S}earch \textbf{G}enerative \textbf{R}etrieval (\textbf{TSGR}), a unified generative retrieval framework that incorporates value awareness into both item representation and candidate ranking. 1) For item representation, TSGR introduces \textbf{Query-aware Parallel SID (QP-SID)}, which encodes query-conditioned value orderings into the SID construction by building parallel codebooks derived from query-item statistics, so that higher-value and query-relevant items are assigned better token indices. 2) For candidate ranking, we introduce a \textbf{Value-aware Ranking Module (VRM)} that is built upon and jointly optimized with the GR, enabling a single model to seamlessly serve as both retriever and pre-ranker without a dedicated pre-ranking stage. A progressive training pipeline further aligns the model with semantic relevance, user preferences, and business objectives.
Offline experiments show that TSGR achieves a 9.16\% improvement in HR@1000, and online A/B tests further validate its effectiveness, yielding gains of +0.43\% in IPV, +1.12\% in Transaction Count, and +1.64\% in GMV. TSGR has been fully deployed in production.

%% file: Sections/1_Introduction.tex
\section{Introduction}

% ===== Paragraph 1 =====
Generative retrieval (GR) has emerged as a promising paradigm for the retrieval stage of industrial e-commerce search, leveraging the sequence generation capability of LLMs to train a single autoregressive model that directly generates the Semantic IDs (SIDs) of target items~\cite{tay2022transformer,rajput2023recommender}.
This formulation enables end-to-end optimization while considerably reducing the infrastructure overhead of maintaining large-scale indices.
Building on these advantages, GR has been rapidly deployed at scale across a range of industrial applications, including recommendation~\cite{onerec2024}, search~\cite{chen2025onesearch}, and advertising~\cite{zhang2025gpr,zhang2026unified}.

% ===== Paragraph 2 =====
Beyond these practical benefits, GR is conceptually well aligned with the nature of information retrieval: the generative model encodes knowledge of the entire item collection into its parameters and reconstructs relevant items in response to a query, while SID encoding compresses the vast product space into a compact token vocabulary, making full-space optimization feasible and alleviating the negative sampling bias in conventional dual-encoder models. Together, these properties make GR a strong foundation for large-scale semantic matching in industrial search.

% ===== Paragraph 3 =====
However, industrial e-commerce search is ultimately driven by business objectives (\eg, CTR, CVR and GMV). Downstream ranking stages rely heavily on these signals to score candidates, precisely because two items equally relevant to a query may differ markedly in conversion probability. When the retrieval stage is unaware of value signals, it may retrieve semantically reasonable but commercially inferior candidates, limiting the business impact of downstream rankers~\cite{zhang2026unified}.
This insensitivity to business value stems from how existing e-commerce GR frameworks operate, where value signals are seldom incorporated into either item representation or model output, giving rise to two key limitations.

\textbf{Value-insensitive SID construction.}
Existing RQ-based SID pipelines are primarily designed to preserve multimodal semantic similarity and generally lack an explicit mechanism to reflect item business value~\cite{rajput2023recommender,onerec2024}.
As a result, items of similar semantics but markedly different value tend to be mapped to the same SID neighborhood, yielding a retrieval space that is semantically organized but largely value-undifferentiated.
Moreover, SIDs are typically fixed globally and remain query-agnostic, so that each item carries an identical identifier regardless of query, overlooking how its relevance and value can vary across different search intents.

\textbf{Value-unaware candidate ranking.}
Standard GR ranks candidates by the probability of SID tokens under beam search, reflecting semantic relevance more than business relevance.
Furthermore, since candidates are generated dynamically at inference time, the model cannot prefetch rich item side-info at input (\eg, features and value signals).
Consequently, the candidates ranked by beam search score often diverge from the items most likely to be clicked or purchased.

% ===== Paragraph 5 =====
To address these limitations, we propose \textbf{T}aobao \textbf{S}earch \textbf{G}enerative \textbf{R}etrieval (\textbf{TSGR}), a unified generative framework that incorporates value awareness into both item representation and candidate ranking via two core innovations.
% For item representation, we design \textbf{Query-aware Parallel SID (QP-SID)}, which encodes value signals directly into the SID space and constructs query-conditioned orderings guided by query-item click statistics, producing identifiers that capture both business value and query-specific relevance.
For item representation, we observe that items sharing the same third-level SID token mutually reinforce each other's training signal, motivating us to assign high-value items to earlier token indices. Building on this insight, we propose \textbf{Query-aware Parallel SID (QP-SID)}, which constructs multiple parallel orderings of items, including a default ordering based on overall clicks and query-conditioned orderings based on query-specific click statistics. By assigning higher-value and query-relevant items to earlier indices, QP-SID naturally aligns generation likelihood with both business value and query relevance.
For candidate ranking, we introduce a \textbf{Value-aware Ranking Module (VRM)} that extracts user representations from the hidden states of the generative backbone and fuses them with rich item side-info via cross-attention, producing scores aligned with business objectives for re-ranking beam search candidates.
Crucially, this module is jointly optimized with the generation objective within the same model, naturally extending GR into a unified retrieval and pre-ranking framework without a dedicated pre-ranking stage.
To further align TSGR with downstream business objectives, we adopt a progressive training pipeline consisting of Pre-SFT multi-task pretraining and SFT with weighted multi-positive supervision. We also explore the role of reinforcement learning (RL), and find that applying the VRM makes superior performance.

% ===== Contributions =====
Our main contributions are summarized as follows:
\begin{itemize}
    \item We propose \textbf{TSGR}, a unified generative framework that integrates retrieval and pre-ranking within a single model and aligns the full pipeline with business objectives.
    \item We design \textbf{QP-SID}, an item representation that encodes value signals into SIDs and derives query-aware orderings from query-item click statistics, addressing the value-insensitivity and query-agnosticism of conventional SIDs.
    \item We introduce the \textbf{VRM} that fuses backbone-derived user representations with item side-info via cross-attention, and jointly optimize it with the generation backbone so that a single model serves as both retriever and pre-ranker.
    \item We conduct extensive offline experiments on real-world Taobao search data and large-scale online A/B tests,  yielding gains of +0.43\% in IPV, +1.12\% in Transaction Count, and +1.64\% in GMV. TSGR has been fully deployed in production.
\end{itemize}

%% file: Sections/2_RelatedWorks.tex
\section{Related Work}

\subsection{Generative Retrieval and Recommendation}

Generative retrieval and recommendation reframe item retrieval as a sequence generation problem, in which a model directly produces the discrete SIDs of items in an autoregressive manner.
DSI~\cite{tay2022transformer} first demonstrates this paradigm by memorizing a document corpus within model parameters.
TIGER~\cite{rajput2023recommender} employs RQ-VAE to compress item embeddings into hierarchical token sequences, unifying semantic and collaborative signals through a shared quantized vocabulary, while IDGenRec~\cite{tan2024idgenrec} further connects generative models with item ID spaces via textual ID generation.

As generative models scale to billion-parameter LLMs, this paradigm has evolved from proof-of-concept to industrial deployment.
On the architectural side, HSTU~\cite{zhai2024hstu} and HLLM~\cite{chen2024hllm} improve efficiency through tailored attention mechanisms and hierarchical LLM designs, whereas OneRec~\cite{onerec2024} casts recommendation as a generic sequence prediction task with preference alignment.
In e-commerce search, OneSearch~\cite{chen2025onesearch} proposes a unified end-to-end generative framework, which OneSearch-V2~\cite{chen2026onesearch} extends with latent reasoning and self-distillation.
Similar trends have also emerged in advertising~\cite{zhang2025gpr,zheng2025ega,zhang2026unified} and general search~\cite{han2025mtgr,hao2025oxygenrec}.

\subsection{Semantic ID Construction}

Semantic IDs convert items into compact discrete token sequences that serve as generation targets for generative retrieval.
TIGER~\cite{rajput2023recommender} pioneers RQ-VAE for hierarchical quantization, and subsequent works simplify this process via RQ-KMeans~\cite{onerec2024}, OPQ~\cite{hou2025rpg}, and joint product quantization~\cite{petrov2024recjpq}.
To enrich SID semantics, LETTER~\cite{wang2024letter} and LC-Rec~\cite{zheng2024lcrec} align multimodal features with collaborative signals, whereas GSID~\cite{yang2025gsid}, CAT-ID2~\cite{liu2026cat}, and Hi-Gen~\cite{wu2024hi} inject domain-specific structures such as attribute hierarchies and category trees.
CQ-SID~\cite{zhu2026efficient} encodes items into semantic clusters rather than unique identifiers to reduce beam search cost.
At industrial scale, FORGE~\cite{fu2025forge} benchmarks SID construction on Taobao with 14 billion interactions, and UniSID~\cite{unisid} jointly optimizes SID construction with the recommendation objective in an end-to-end manner.

All existing methods assign each item a single global SID, confining it to a fixed representation regardless of query context and overlooking item business value during codebook construction.
TSGR addresses both limitations by constructing parallel SIDs per item via query-aware, value-informed codebooks, allowing the same item to be associated with different representations across queries while prioritizing high-value items within each cluster.

\subsection{Value-aware Ranking}

Industrial search systems typically adopt a cascaded architecture in which retrieval and ranking are handled by separate models~\cite{pei2019value}, which tends to introduce objective misalignment and prevents item value signals from participating in candidate generation~\cite{wang2026towards}.
To narrow this gap, a growing line of work explores using generative models directly for ranking.
GenRank~\cite{huang2025towards} demonstrates generative ranking at industrial scale, and QGS~\cite{song2026item} integrates heterogeneous features into generative ranking through HFG-Attention.
UniVA~\cite{zhang2026unified} introduces a Generation-as-Ranking formulation with eCPM-aware reinforcement learning, while GoalRank~\cite{zhang2025goalrank}, GReF~\cite{lin2025gref}, and DeGRe~\cite{song2026degre} incorporate ranking signals through reward models and dense supervision.

However, these methods largely rely on generation probabilities as ranking scores, which reflect sequence likelihood rather than actual item value, offering limited support for the rich item side-info that are critical for value estimation in industrial systems. In contrast, TSGR directly incorporates item side-info into value-aware scoring within the generative model, unifying retrieval and ranking under a single jointly optimized objective.

%% file: Sections/3_Methodology.tex
\section{Methodology}
\label{sec:method}

\begin{figure*}
    \centering
    \includegraphics[width=0.98\linewidth]{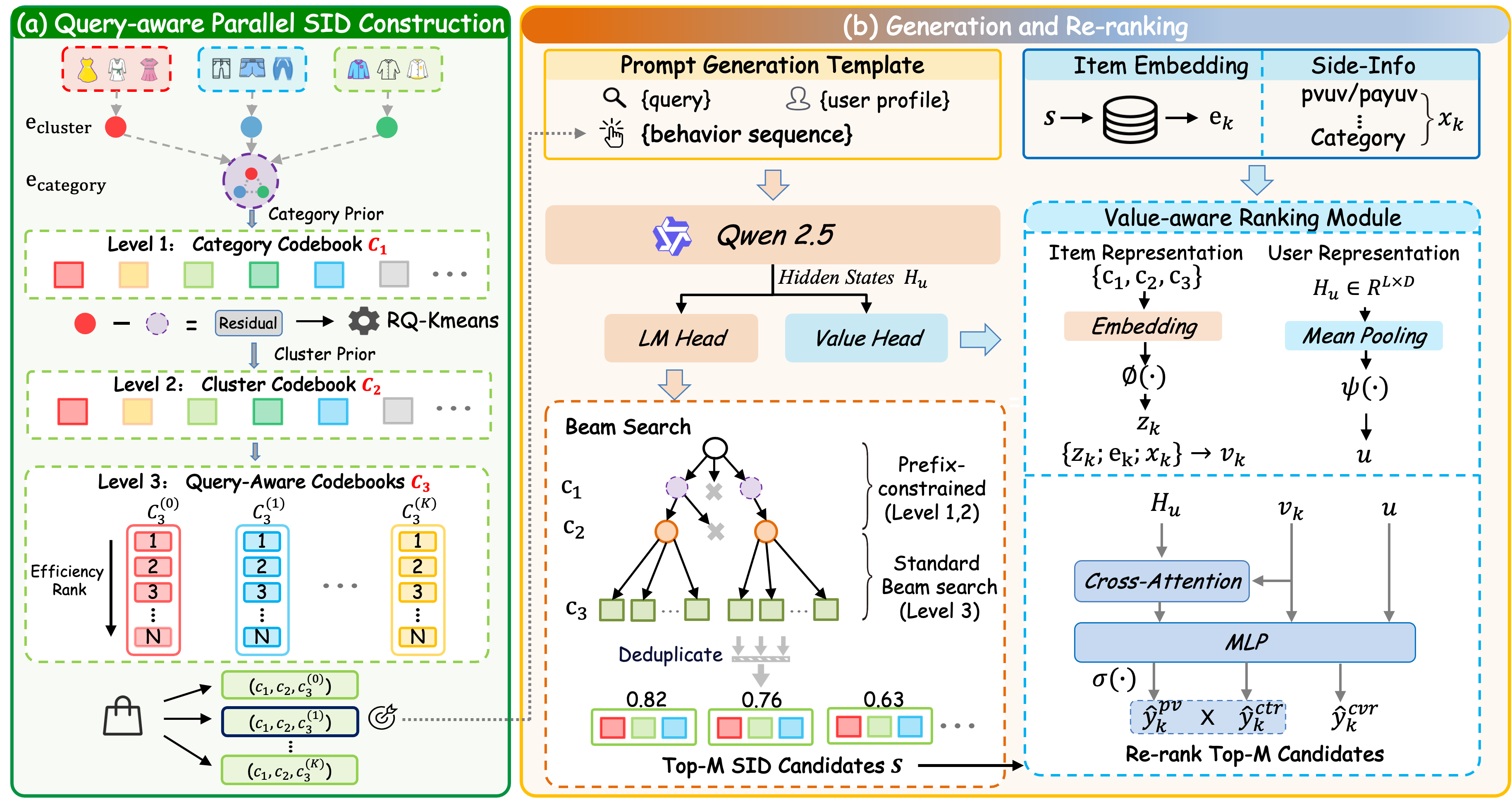}
    \caption{Overview of TSGR, a unified generative framework that jointly handles retrieval and pre-ranking within a single model.
    \textbf{(a) Parallel SID Construction} builds a three-level codebook using category and cluster priors at Levels~1--2, and query-aware codebooks at Level~3.
    \textbf{(b) Generative Backbone} performs hierarchical beam search with prefix-constrained decoding at Levels~1--2 and standard beam search at Level~3.
    \textbf{VRM} re-ranks candidates by fusing SID embeddings with item side-info.}
    \label{fig:main}
\end{figure*}

TSGR is a unified generative framework for industrial e-commerce search that integrates retrieval and pre-ranking within a single model.
As illustrated in Figure~\ref{fig:main}, the \textbf{Query-aware Parallel SID Construction} (Section~\ref{sec:parallel_sid}) builds distinct SID paths for each item and selects the most appropriate one conditioned on the query.
The \textbf{Value-aware Ranking Module} (Section~\ref{sec:value_aware}) is built upon the generative backbone and re-ranks beam search candidates.
Section~\ref{sec:training_inference} presents a training pipeline that progressively develops model capability through Pre-SFT multi-task pretraining and SFT with weighted multi-positive training. Besides, We also explore the role of RL.

\subsection{Query-aware Parallel SID Construction}
\label{sec:parallel_sid}

We propose a parallel SID construction method that builds distinct SID representations for each item based on query-item statistics. Specifically, we leverage two types of prior information to construct two codebooks: a semantic codebook configured as $32768 \times 8192$ to capture the semantic structure of items, and an efficiency codebook of overall size $8192$ composed of $N$ parallel branches, enabling diverse retrieval paths while concentrating the decoding budget on high-probability branches and avoiding unnecessary search capacity on low-probability paths~\cite{CRSID}.

\noindent\textbf{Semantic Codebook Construction.}
The first two levels form the semantic codebook, which is built upon existing item features such as same-item cluster and item category, to construct a natural item hierarchy \textit{Item $\in$ Cluster $\in$ Category}. 
We first obtain cluster-level and category-level embeddings by averaging the embeddings of all items within each group:

\begin{equation}
    \mathbf{e}_{\text{cat}}(g) = \frac{1}{|\mathcal{G}|}\sum_{i \in \mathcal{G}} \mathbf{e}_i \qquad
    \mathbf{e}_{\text{cluster}}(g) = \frac{1}{|\mathcal{C}|}\sum_{i \in \mathcal{C}} \mathbf{e}_i
\end{equation}
where $\mathcal{G}$ and $\mathcal{C}$ denote items in the same category and cluster, and $\mathbf{e}_i \in \mathbb{R}^d$ is the embedding of item $i$.
The Level-1 codebook is initialized directly with category embeddings, as the number of categories ($8100+$) closely matches the codebook size of $8192$. RQ-KMeans is then applied to the residuals between cluster and category embeddings to learn the Level-2 codebook:
\begin{equation}
    \mathbf{r}(g) = \mathbf{e}_{\text{cluster}}(g) - \mathbf{e}_{\text{cat}}(g)
\end{equation}
This produces a three-level $8192 \times 4 \times 8192$ intermediate codebook. Since a four-level codebook in total would incur prohibitive deployment cost in production, we merge the first two levels into a single level, resulting in the final semantic codebook of size $32768 \times 8192$.

This construction injects two structural priors into the semantic codebook:
\begin{itemize}
    \item \textit{Category prior}: the domain taxonomy is embedded directly into the coarsest partition, so that the SID space inherits a meaningful category structure.
    \item \textit{Cluster prior}: by quantizing cluster centroids rather than individual item embeddings, all items within the same cluster are guaranteed to share an identical semantic SID prefix.
\end{itemize}
Together, these priors make the SID space semantically well-organized and provide a clean cluster partition on which the efficiency codebook is built.

\noindent\textbf{Efficiency Codebook Construction.}
The third level constitutes the efficiency codebook, constructed independently within each cluster by ordering items according to their click-based value. Higher-value items occupy earlier token indices, which are visited more frequently during training, concentrating the decoding budget on the most promising items and aligning generation likelihood with item value.

A straightforward design ranks items by their click count over the past 30 days, giving the default ordering:
\begin{equation}
    \pi_0 = \operatorname{argsort}_{i \in \mathcal{C}}\, s_0(i)
    \label{eq:default_ordering}
\end{equation}
where $\mathcal{C}$ denotes the set of items in the cluster and $s_0(i)$ is the click count of item $i$. However, this ordering remains query-agnostic, overlooking the fact that an item may be highly relevant to one query but not another. To overcome this, we propose \textbf{QP-SID}, built on the observation that each query can be segmented into key terms and each item belongs to one cluster, so term-cluster relations can be inferred from query-item statistics. Concretely, for each cluster we identify the top-3 most representative terms, filtering out those whose coverage exceeds $80\%$, where coverage is defined as the fraction of items in the cluster clicked under the term:

\begin{equation}
    \text{cov}(t, \mathcal{C}) = \frac{|\{i \in \mathcal{C} : s(i, t) > 0\}|}{|\mathcal{C}|}
    \label{eq:coverage}
\end{equation}
A term with near-complete cluster coverage carries little discriminative signal, as it is associated with virtually all items regardless of query intent. For each retained representative term $t$, let $s(i, t)$ denote the click count of item $i$ under term $t$ over the past 30 days. Ranking items by $s(i, t)$ produces the term-specific ordering:
\begin{equation}
    \pi_t = \operatorname{argsort}_{i \in \mathcal{C}}\, s(i, t)
    \label{eq:term_ordering}
\end{equation}
This produces $3$ term-conditioned orderings together with the default ordering $\pi_0$, giving $4$ parallel orderings per cluster. The case study can be found in Appendix~\ref{app:QPSID}.

\input{Tables/presft}
\subsection{Value-aware Ranking Module}
\label{sec:value_aware}
In GR, candidates are ranked by generation probability, which reflects sequence likelihood rather than business value. Since candidate items are only known after decoding, their side-info cannot be incorporated during generation. A natural remedy is a separate downstream pre-ranking model, but this couples two models with misaligned objectives: the GR model optimizes SID generation likelihood over its own retrieval distribution, while the pre-ranking model scores candidates from the full multi-source pool, introducing a distribution gap that undermines their combination.

To this end, we introduce the \textbf{VRM} on top of the generative backbone.
As illustrated in Figure~\ref{fig:main}(b), the module takes the user representations of the generative model together with item side-info as input, and outputs a value score for each candidate.
We describe its four components in turn: item representation, user representation, user-item interaction and candidates scoring.

\paragraph{Item Representation.}
For each candidate $(c_1, c_2, c_3) \in \mathcal{S}$ from beam search, we construct the item representation from three complementary sources.
The first is an item side-info vector $\mathbf{x}_k \in \mathbb{R}^{d_f}$ retrieved from offline feature tables, which aggregates four categories of features:
(1)~behavioral statistics over multiple time windows (\eg, page views and payment counts);
(2)~category features derived from the item's category hierarchy;
(3)~global seller statistics that capture cross-scenario behavioral patterns; and
(4)~query-item co-occurrence features.
The second is an item embedding $\mathbf{e}_k$ retrieved from the offline feature table, which provides information that may be lost during SID construction.
The third is a SID embedding, obtained by mapping the candidate SIDs $(c_1, c_2, c_3)$ through the backbone's embedding layer and compressing it with an MLP $\phi(\cdot)$:
\begin{equation}
    \mathbf{z}_k = \phi\big(\text{Emb}(c_1),\ \text{Emb}(c_2),\ \text{Emb}(c_3)\big)
\end{equation}
The three are concatenated to form the final item representation:
\begin{equation}
    \mathbf{v}_k = [\mathbf{x}_k;\ \mathbf{e}_k;\ \mathbf{z}_k]
\end{equation}

\paragraph{User Representation.}
The user representation reuses the hidden states of the generative backbone, thus incurring no additional encoding cost.
Given the input prompt of length $L$, the backbone produces contextualized representations $\mathbf{H}_u \in \mathbb{R}^{L \times D}$, where $D$ is the hidden dimension.
A global user representation is obtained by mean pooling over all token positions followed by an MLP $\psi(\cdot)$:
\begin{equation}
    \mathbf{u} = \psi\!\left(\frac{1}{L}\sum_{i=1}^{L} \mathbf{H}_u[i]\right)
\end{equation}

\paragraph{User-Item Interaction.}
To capture fine-grained relevance beyond the global summary, we introduce an attention module in which the item representation $\mathbf{v}_k$ attends to the user context representations $\mathbf{H}_u$:
\begin{equation}
    \mathbf{a}_k = \text{CrossAttn}(\mathbf{v}_k,\ \mathbf{H}_u,\ \mathbf{H}_u)
\end{equation}
where $\mathbf{v}_k$ serves as the query and $\mathbf{H}_u$ serves as both the key and the value.
This allows each candidate to selectively attend to the most relevant parts of the user context, yielding an interaction representation $\mathbf{a}_k$ that encodes user-item relevance.

\paragraph{Feature Fusion and Scoring.}
The interaction representation, the user representation, and the item representation are concatenated and fed into three parallel heads to produce corresponding prediction scores:
\begin{equation}
    \hat{y}_k^{t} = \sigma\big(\text{MLP}_{t}\big([\mathbf{a}_k;\ \mathbf{u};\ \mathbf{v}_k]\big)\big) \quad t \in \{\text{pv},\ \text{ctr},\ \text{cvr}\}
\end{equation}
where $\sigma(\cdot)$ is the sigmoid function. At inference, candidates are re-ranked by the composite score $\hat{y}_k^{\text{pv}} \times \hat{y}_k^{\text{ctr}}$ to produce the final output.
Since this module shares the backbone embedding and reuses the hidden states computed during generation, it unifies retrieval and value-aware ranking within a single model at negligible additional cost.

\input{Tables/main_table}
\subsection{Training Pipeline}
\label{sec:training_inference}
\begin{figure}
    \centering
    \includegraphics[width=0.98\linewidth]{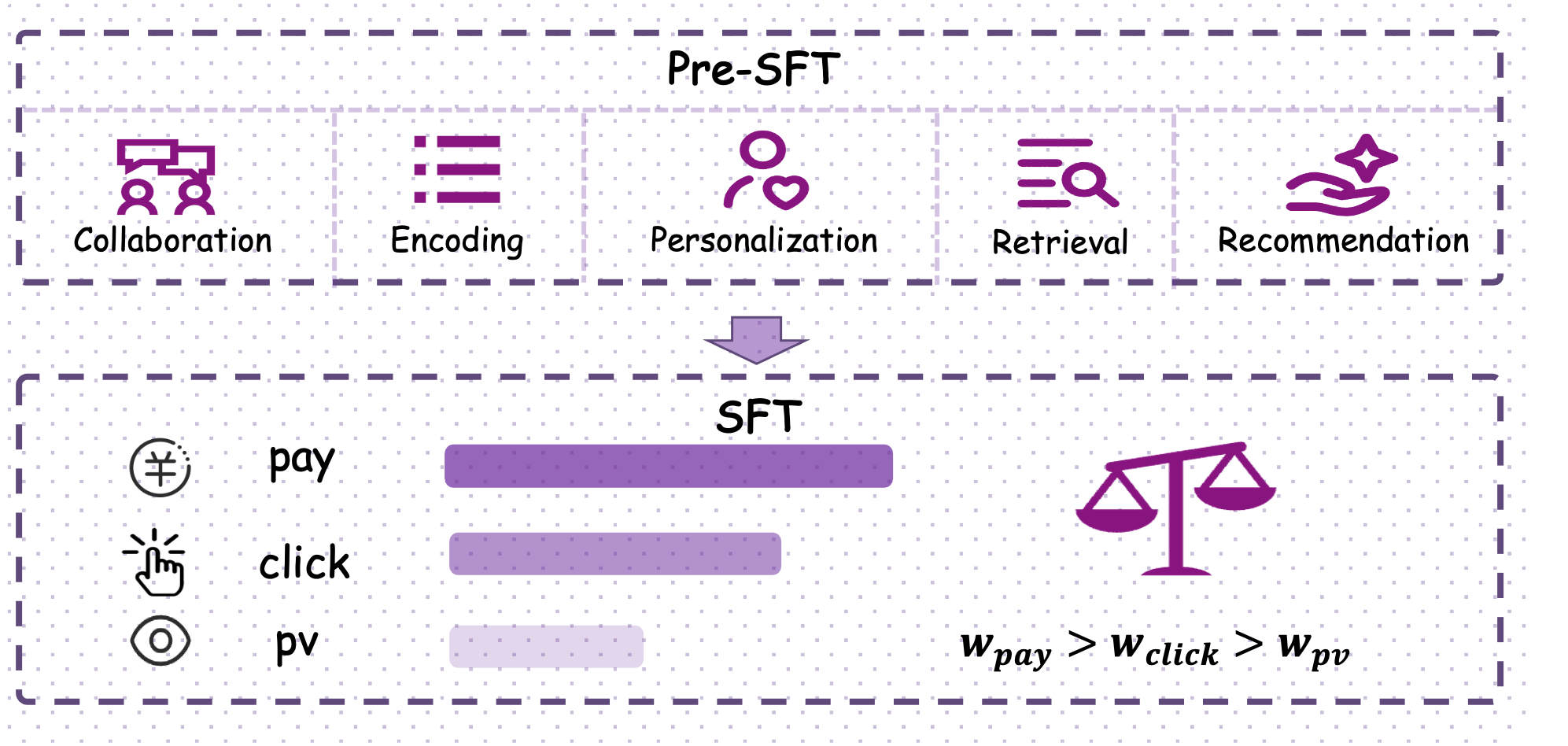}
    \caption{Training Pipeline. Model capabilities are developed progressively via Pre-SFT multi-task pretraining and SFT with a weighted multi-positive loss.}
    \label{fig:training}
\end{figure}
Training TSGR requires three capabilities to be developed together: semantic alignment between item content and the SID space, personalized query-to-SID generation, and value-aware candidate ranking.
Acquiring all three within a single supervised stage is difficult, so we adopt a pipeline that builds them progressively as shown in Figure~\ref{fig:training}, with detailed prompt templates provided in Appendix~\ref{app:prompts}.

\subsubsection{Pre-SFT}
\label{sec:presft}
Pretrained LLMs are not equipped to map items or queries to discrete SID tokens.
Instead of forcing the model to acquire this and all other capabilities at once, Pre-SFT decomposes the target objective into simpler sub-tasks, each focusing on a single capability dimension.
Training on large-scale samples across these sub-tasks activates the necessary capabilities in advance, which lowers both the learning difficulty and the data requirements of the subsequent SFT stage.

We accordingly design five categories of Pre-SFT tasks, summarized in Table~\ref{tab:presft_tasks}.
Most tasks share a unified SID output format but differ in input context.
During training, all tasks are shuffled and optimized jointly with the standard cross-entropy loss.

\subsubsection{SFT}
\label{sec:sft}

The SFT stage adapts the Pre-SFT-initialized model to the personalized retrieval task.

\paragraph{Prompt Generation Framework.}
To support flexible feature iteration, we build the SFT pipeline on top of \textit{Prompt Generation} (PG)~\cite{ou2026promptgenerationtechnicalreport}, which offers a unified prompt template with configurable feature slots.
This enables rapid feature experiments while ensuring training-serving consistency.

\paragraph{Data Construction.}
To construct data with QP-SID, the third-level token is jointly determined by the query $q$ and the cluster of item $i$. Given a query $q$, we extract its key terms and match them against the representative terms $\mathcal{T}_{\text{cluster}}$ of item $i$'s cluster. When a match exists, we select the most representative matched term:
\begin{equation}
    t^* = \operatorname{arg\,max}_{t \,\in\, \mathcal{T}_{\text{cluster}} \cap \operatorname{terms}(q)} \operatorname{rel}(t, \mathcal{C})
\end{equation}
where $\operatorname{rel}(t, \mathcal{C})$ is the co-occurrence click count between term $t$ and cluster $\mathcal{C}$. The third-level token is then assigned as:
\begin{equation}
c_3(i, q) =
\begin{cases}
\pi_{t^*}(i) & \text{if } \mathcal{T}_{\text{cluster}} \cap \operatorname{terms}(q) \neq \emptyset \\
\pi_0(i) & \text{otherwise}
\end{cases}
\label{eq:c3}
\end{equation}
making the third-level token a dynamic function of both the item and the query: an item is assigned an earlier index within its cluster when the query closely aligns with its dominant usage context, producing query-aware training labels that reflect fine-grained item-query relevance.

\paragraph{Session-wise Weighted Multi-Positive Training.}
For each search session, items from four engagement levels (pay, click, pv, and unpv) are packed into a single training sample, with interacted items (pay, click, pv) serving as multiple labels. Attention masks are applied between labels to prevent cross-label interference. Each label is assigned a priority weight satisfying $w_{\text{pay}} > w_{\text{clk}} > w_{\text{pv}}$. The generation loss is a weighted cross-entropy over all valid labels:
\begin{equation}
\label{eq:loss}
    \mathcal{L}_{\text{gen}} = \sum_{k \in \mathcal{Y}} w_k \cdot \text{CE}\big(\mathbf{o}_k,\ t_k\big)
\end{equation}
where $\mathcal{Y}$ is the set of interacted items in the session, $\mathbf{o}_k$ are the logits at the decoding positions of item $k$'s SID tokens, $t_k$ is the ground-truth SID token, and $w_k$ is the corresponding weight.

The VRM is trained jointly in this stage.
Its ranking loss consists of three binary cross-entropy terms over the predicted scores for PV, CTR, and CVR, defined on the same $\mathcal{Y}$:
\begin{equation}
    \mathcal{L}_{\text{rank}} = \sum_{k \in \mathcal{Y}} \sum_{s \in \{\text{pv, ctr, cvr}\}} \text{BCE}\big(\hat{y}_k^{s},\ y_k^{s}\big)
\end{equation}

where $\hat{y}_k^t$ is the predicted score and $y_k^t$ is the corresponding label for task $t$. The overall objective combines the two terms:
\begin{equation}
    \mathcal{L} = \mathcal{L}_{\text{gen}} + \lambda \cdot \mathcal{L}_{\text{rank}}
\end{equation}

Vanilla SFT aligns the model with observed user behavior, but falls short of directly optimizing downstream user satisfaction signals. To push performance further, we build a complete RL pipeline based on GRPO~\cite{shao2024deepseekmath}, using the online ranking model as the reward source to provide direct business-objective feedback. However, compared to introducing item side-info via the ranking module (Section~\ref{sec:value_aware}), the RL-based approach makes limited additional gains. We therefore adopt the ranking module as our primary solution, with full RL experimental details provided in Appendix~\ref{app:rl}.

% \subsubsection{RL}
% \label{sec:rl}

% Vanilla SFT aligns the model with observed user behavior, yet falls short of directly optimizing the downstream signals that reflect actual user satisfaction. To push performance further beyond SFT, one approach is to explicitly introduce item side-info for re-ranking (Section~\ref{sec:value_aware}). Alternatively, incorporating an RL stage based on GRPO~\cite{shao2024deepseekmath} allows the model to internalize actual user satisfaction signals through reward feedback. Although we ultimately adopt the former, we share our RL exploration here for completeness, with implementation details provided in Appendix~\ref{app:rl}.

% \subsubsection{Inference}
% \label{sec:inference}

% TSGR is deployed as a unified service that replaces the conventional separate retrieval and pre-ranking stages, with its outputs passed directly to the ranking stage.
% Given a user query and context features, the model performs constrained beam search.
% The first two SID levels are decoded under a valid-prefix constraint enforced by a pre-built prefix tree, which prunes invalid token sequences, while the third level uses standard beam search, yielding a top-$N$ set of complete SID sequences.
% As each candidate is generated, the value-aware ranking module computes a value score.
% The candidates are then re-ranked and returned as the final output.

%% file: Tables/presft.tex
\begin{table*}[t]
\caption{Pre-SFT task design. Each task activates a specific capability dimension by varying the input context. Most tasks share a unified SID output format; the title generation task uses natural language output to activate intent understanding at the textual level.}
\label{tab:presft_tasks}
\centering
\small
\begin{tabular}{@{}llll@{}}
\toprule
\textbf{Task Category} & \textbf{Input} & \textbf{Output} & \textbf{Capability} \\
\midrule
Encoding & Item title & SID & Learn item semantics to SID space mapping \\
Personalization & User profile + query + behavior seq. & Title & Activate user intent understanding at textual level \\
Retrieval & Search query & SID & Establish query--item relevance for in-domain retrieval \\
Collaboration & Trigger item (title + SID) & SID & Inject co-purchase collaborative signals into SID space \\
Recommendation & User profile + behavior seq. & SID & Enable personalization and cross-domain generalization \\
\bottomrule
\end{tabular}
\end{table*}

%% file: Tables/main_table.tex
\definecolor{lightyellow}{RGB}{255,255,204}

\begin{table*}[t]
\centering
\caption{Performance comparison of the proposed methods on the Taobao search dataset. Within each category, methods are mutually exclusive and $^\dagger$ marks the methods adopted in TSGR. Best results are in \textbf{bold}.}
\label{tab:main}
\resizebox{0.85\textwidth}{!}{
\begin{tabular}{l l ccccc}
\toprule
\textbf{Category} & \textbf{Method}
& \textbf{HR@20} & \textbf{HR@100} & \textbf{HR@500} & \cellcolor{lightyellow}\textbf{HR@1000} & \textbf{HR@5000} \\
\midrule
\multirow{2}{*}{SID}
& FORGE
& 0.4328 & 0.5863 & 0.7207 & \cellcolor{lightyellow}0.7735 & 0.8604 \\
& QP-SID$^\dagger$
& 0.4635 & 0.6235 & 0.7669 & \cellcolor{lightyellow}0.8177 & 0.8961 \\
\midrule
Pre-processing & Pre-SFT$^\dagger$
& \textbf{0.4694} & 0.6340 & 0.7740 & \cellcolor{lightyellow}0.8222 & 0.8987 \\
\midrule
\multirow{3}{*}{Post-processing}
& Pre-Ranking Model
& 0.4248 & 0.6264 & 0.8112 & \cellcolor{lightyellow}0.8552 & 0.8771 \\
& RL
& 0.4659 & 0.6420 & 0.7857 & \cellcolor{lightyellow}0.8277 & \textbf{0.9027} \\
& \textbf{VRM (TSGR)}$^\dagger$
& 0.4586 & \textbf{0.6677} & \textbf{0.8250} & \cellcolor{lightyellow}\textbf{0.8651} & 0.8953 \\
\bottomrule
\end{tabular}
}
\end{table*}

%% file: Sections/4_Experiment.tex
\section{Experiments}
\label{sec:experiments}
To more thoroughly assess the effectiveness of TSGR, in this section we conduct comprehensive offline and online A/B evaluations. Moreover, extensive ablation experiments are performed to prove the feasibility of each module.
\subsection{Experimental Setup}
\label{sec:exp_setup}

\paragraph{Dataset and Baseline.}
We collect over 200 million real user interactions from Taobao's e-commerce search platform over two weeks as training data, with the test set sampled from the following day. To assess generalizability of the VRM, we additionally conduct ablation experiments on it using the TencentGR~\cite{pan2026tencent} dataset in Appendix~\ref{app:tencent_gr}. We compare against two baselines: \textbf{FORGE}~\cite{fu2025forge}, which follows its approach by randomly initializing the last-level token as the SID baseline, where each corresponds to up to 5 items;
and \textbf{Pre-Ranking Model}, a conventional pre-ranking model applied after retrieval.

\paragraph{Evaluation Metrics.}
To verify retrieval performance, we adopt HitRate@$K$ (HR@$K$) as evaluation metric. Following standard practice in industrial generative retrieval~\cite{zhu2026efficient,chen2025onesearch}, we report HR@$K$ under the same beam size with \colorbox{lightyellow}{\strut HR@1000} as the primary metric, as the goal of retrieval is to ensure that relevant items are surfaced within the candidate pool passed to downstream rankers.

\paragraph{Implementation Details.}
We adopt Qwen2.5-0.5B~\cite{yang2024qwen2} as the generative backbone, with a three-level SID codebook ($32768{\times}8192{\times}8192$), beam sizes of $[400, 400, 10000]$, and $4$ parallel third-level orderings per item from QP-SID. SFT is trained with AdamW ($\beta_1{=}0.9$, $\beta_2{=}0.999$), a cosine learning rate of $1{\times}10^{-4}$, and a batch size of 256, with $\lambda{=}1$ and value score weights $(w_{\text{pv}}, w_{\text{clk}}, w_{\text{pay}}) = (1, 2, 3)$. More details can be found in Appendix~\ref{app:details}.

\subsection{Overall Comparison}
\label{sec:overall}
Table~\ref{tab:main} reports HR@$K$ across all compared methods. We organize experiments around three dimensions: SID construction, pre-processing, and post-processing.

QP-SID outperforms FORGE across all metrics with $+4.42\%$ in HR@1000 by incorporating item value signals and query-conditioned orderings into the efficiency codebook. Within each cluster, higher-value items are assigned earlier token indices, which aggregate more items across clusters and thus receive more training signal. Beyond global value ordering, query-conditioned ordering further promotes query-relevant items to earlier indices, helping the model surface relevant candidates more effectively. Together, these two mechanisms ensure that the model prioritizes both globally high-value and query-specifically relevant items.

The Pre-SFT multi-task pretraining stage further raises performance, demonstrating that this stage contributes substantial gains.

As a baseline for the post-processing stage, the conventional pre-ranking model delivers a notable improvement in HR@500 and HR@1000 ($+3.30\%$) but incurs a performance loss in others. RL training, in contrast, achieves consistent improvements in model capability across nearly all metrics ($+0.55\%$ in HR@1000). However, the magnitude of this improvement remains marginal, so RL has not been adopted in the actual deployment. The VRM achieves stronger performance instead by fusing backbone-derived user representations with item side-info under joint optimization with the generation objective, thereby avoiding cross-stage objective misalignment while introducing no additional serving latency. This module attains an improvement in HR@1000 comparable to that of the pre-ranking model ($+3.73\%$ vs.\ $+3.30\%$), while outperforming it on nearly all others. Together, these results confirm that the unified approach of TSGR to retrieval and pre-ranking is both more effective and more deployment-efficient than the conventional cascaded design.

\subsection{Ablation Study}
\label{sec:ablation}

To better examine the contribution of each design, we evaluate 1)~the impact of SID construction strategies, 2)~the effectiveness of the VRM, and 3)~the key design choices within each training stage. For efficiency, part of the ablation studies are conducted on a sampled subset of the full training data. 

\subsubsection{\textbf{SID Design}}
\begin{figure}
    \centering
    \includegraphics[width=0.98\linewidth]{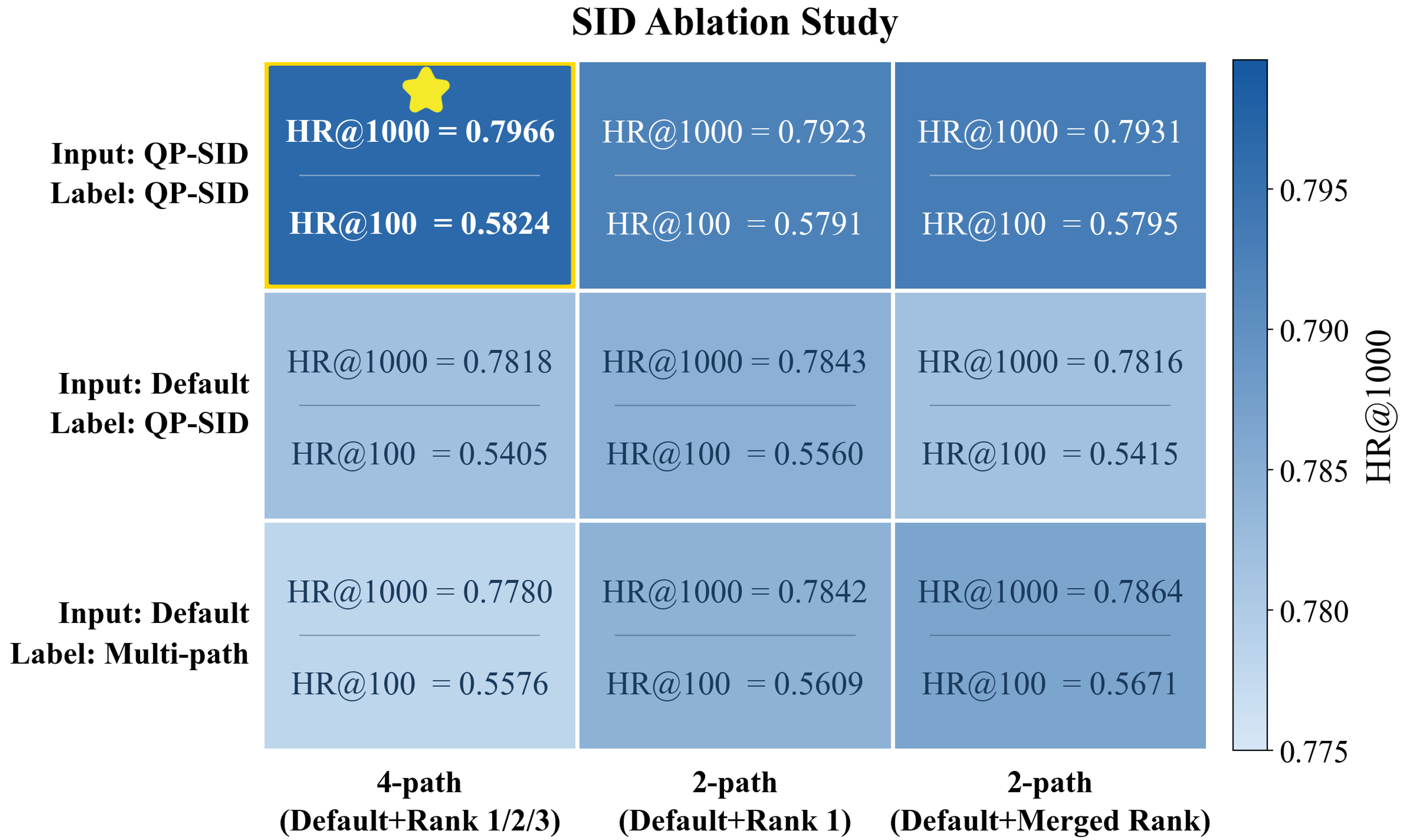}
    \caption{Comparison of different QP-SID configurations and training strategies. HR@1000 and HR@100 are reported for each combination of input/label configuration (rows) and efficiency ID strategy (columns).}
    \label{fig:sid}
\end{figure}
% Table~\ref{tab:ablation_sid} details the evolution of the SID design through a series of progressive improvements, beginning with FORGE and culminating in QP-SID. An intuitive initial step is to incorporate an item efficiency score via re-ranking, which brings noticeable performance gains. However, encoding efficiency information directly into the SID through a dedicated Efficiency ID proves substantially more effective, improving HR@1000 by $+27.51\%$. Furthermore, the semantic ID of FORGE is generated iteratively by an LLM, which restricts the incorporation of explicit item features. To address this limitation, we reconstruct the semantic ID using RQ-Kmeans constrained by category and cluster priors (Section~\ref{sec:parallel_sid}). This reconstruction yields CRID, which matches the performance of the FORGE counterpart while offering greater interpretability and extensibility. Because the resulting four-level codebook introduces deployment challenges and accumulates prediction errors across levels, we consolidate it into a three-level codebook. This modification produces a clear performance gain of $+5.06\%$, indicating that a shallower SID is more robust to error accumulation. Finally, QP-SID replaces the single global efficiency ordering with query-aware parallel orderings. By selecting the ordering that best matches the dominant intent of a given query during serving, this approach achieves further improvements, reaching a final HR@1000 of 0.7966.
We progressively improve the SID design from FORGE to QP-SID. Re-ranking FORGE candidates with an efficiency score yields noticeable gains, but encoding efficiency directly into the SID via a dedicated codebook proves substantially more effective. We further reconstruct the semantic codebook with RQ-KMeans constrained by category and cluster priors (section ~\ref{sec:parallel_sid}), improving interpretability and extensibility. Consolidating the four-level codebook into three levels improves robustness against error accumulation, as a shallower SID is less susceptible to cascading prediction errors. Finally, QP-SID introduces query-aware parallel orderings to replace the single global efficiency ordering, selecting the best-matched ordering per query during serving.

\input{Tables/appendix_qpsid_hit}
\input{Tables/appendix_qpsid_ratio}

Figure~\ref{fig:sid} investigates the impact of different QP-SID configurations and training strategies, revealing two key findings. First, consistency between input and output SID orderings is essential: using QP-SID orderings as training labels while retaining the default ordering as input weakens the model's ability to copy the input SID sequence, causing a substantial HR@100 drop. Second, introducing auxiliary query-matched labels renders the third SID level harder to learn, amplifying its gradient and indirectly undermining the accuracy of the first two levels, as further evidenced by the hierarchical prediction accuracy in Table~\ref{tab:multipath_analysis}. Moreover, we observe that 99.8\% of model outputs collapse to the Default path, indicating that the model fails to leverage the diversity of QP-SID orderings. These observations motivate our final QP-SID design.

Table~\ref{tab:qpsid_distribution} further validates it from a distributional perspective. The majority of items in behavior sequences fall into the Default category, indicating that most historically clicked items are not closely related to the current query. In contrast, up to 76.8\% of training labels are query-relevant (Term Rank = 1, 2, or 3), which is expected since users click with the current query intent in mind. The model output distribution further confirms that the model has successfully learned to retrieve from multiple ranking channels rather than relying solely on the Default channel.

\subsubsection{\textbf{Value-aware Ranking Module}}
\begin{figure}
    \centering
    \includegraphics[width=0.8\linewidth]{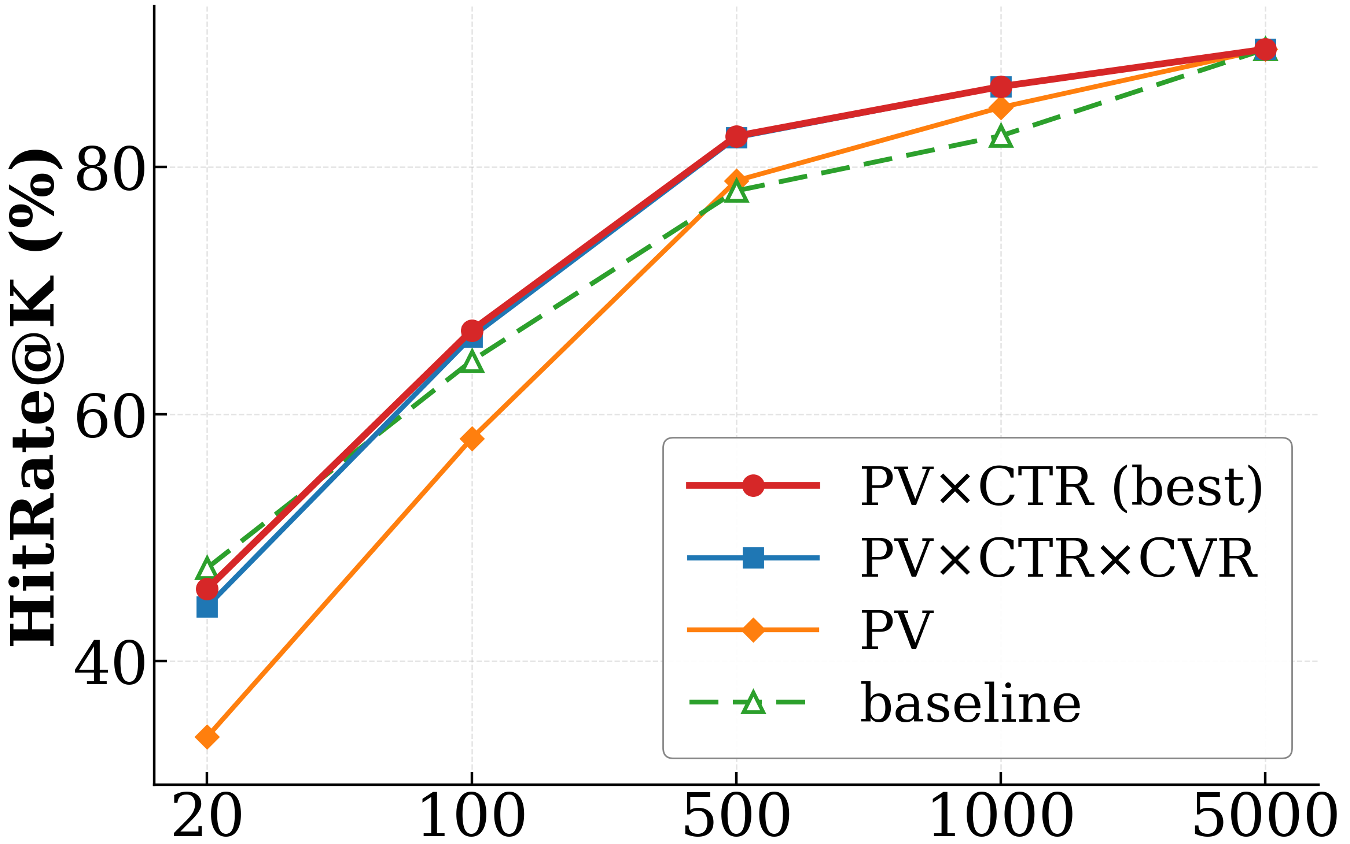}
    \caption{Ablation study on the effect of different score combinations in the VRM.}
    \label{fig:ablation_value_scores}
\end{figure}

\begin{figure}
    \centering
    \includegraphics[width=0.8\linewidth]{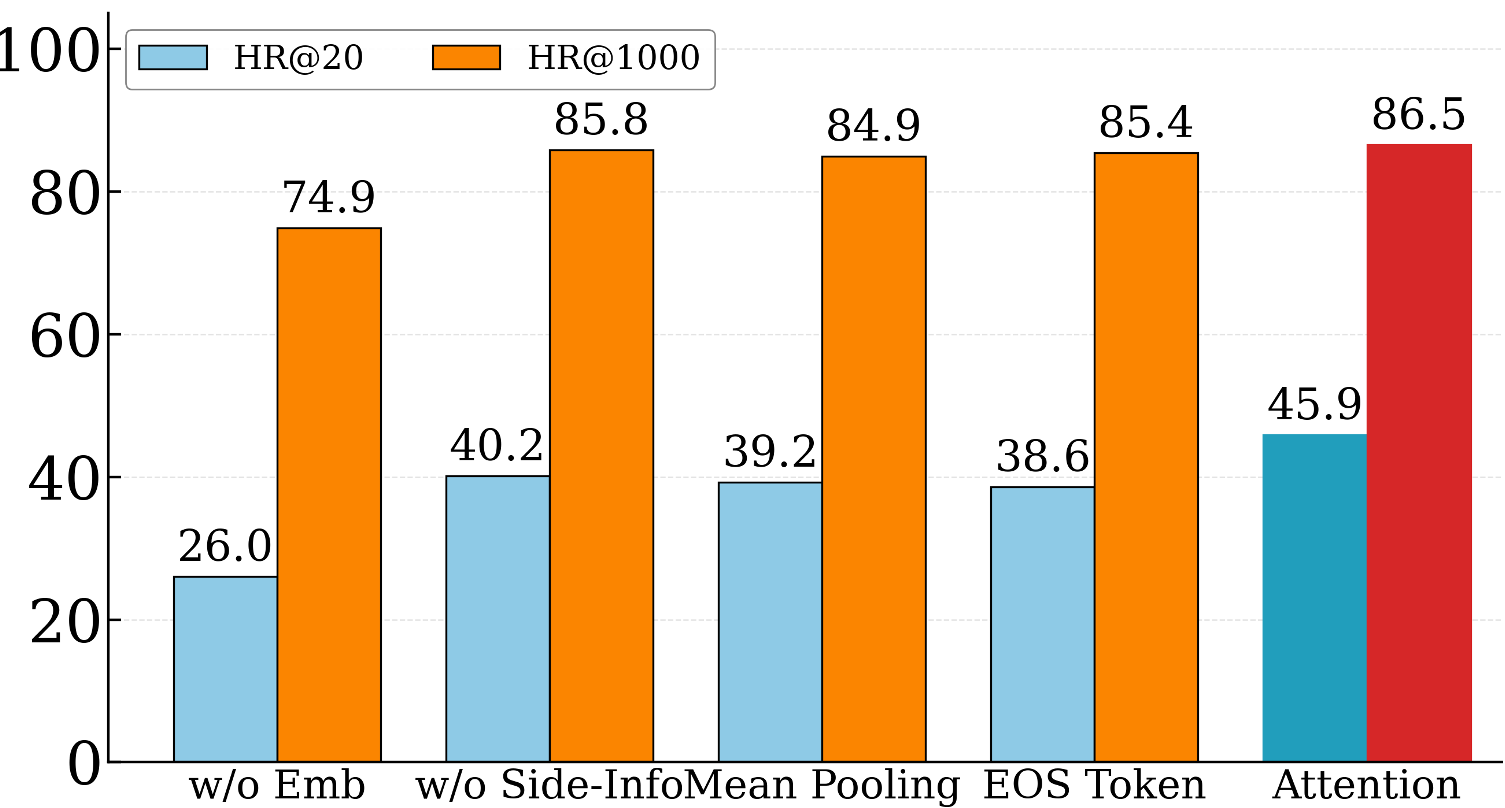}
    \caption{Ablation study on the effect of different architecture variants in the VRM.}
    \label{fig:ablation_value_arch}
\end{figure}
All experiments are conducted on a candidate pool of 5000 items. For the ranking score (Figure~\ref{fig:ablation_value_scores}), we ablate different score combinations used for re-ranking at inference time. All combinations achieve consistent gains over the baseline on HR@1000, and $\text{PV} \times \text{CTR}$ attains the best overall performance. Given that $\text{PV} \times \text{CTR}$ directly corresponds to the HitRate objective, this result is expected. Motivated by this observation, we further examine whether removing the CVR loss from training affects performance, and find that it has negligible impact in our setting.
For the module architecture (Figure~\ref{fig:ablation_value_arch}), removing item embeddings entirely causes severe degradation in both HR@20 and HR@1000, confirming that item representations are indispensable. This is because the SID construction process inevitably discards a substantial amount of item information, making item embeddings a critical complement for recovering the lost semantic detail. Excluding item side-info leads to a substantial drop in HR@20, with only moderate impact on HR@1000. Notably, both replacing the cross-attention module with mean pooling and substituting it with the EOS token representation directly passed into the MLP make a considerable decline in HR@20 while incurring only a marginal drop in HR@1000. This pattern suggests that these two designs can serve as viable lightweight alternatives in latency-sensitive scenarios. More comprehensive ablation results are provided in Appendix~\ref{app:exp}.

\subsubsection{\textbf{Training Strategy}}
% \paragraph{Pre-SFT}
\begin{figure}
    \centering
\includegraphics[width=0.90\linewidth]{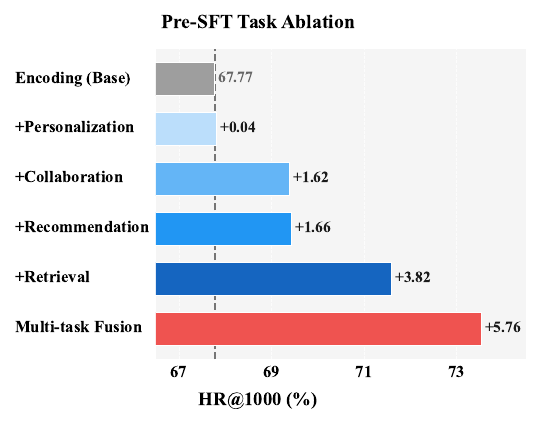}
    \caption{Ablation study on Pre-SFT task design. Each variant adds a single task on top of the Encoding base, while
    Multi-task Fusion trains all tasks jointly.}
    \label{fig:presft_ablation}
\end{figure}
Figure~\ref{fig:presft_ablation} examines the contribution of each Pre-SFT task. Every task yields a positive gain, with the Retrieval task contributing the most due to its closest alignment with the downstream objective. The Personalization task contributes the least, as its title-generation format introduces an output-space gap relative to the SID target. Jointly training on all tasks achieves $+5.76\%$, surpassing any individual task and confirming that the tasks activate complementary capability dimensions.

%   Pre-SFT：有无 Pre-SFT 对比；各类任务的贡献
% \paragraph{SFT}
\input{Tables/sft_ablation}
As shown in Table~\ref{tab:sft_ablation}, the proposed session-wise weighted multi-positive objective consistently outperforms the pointwise baseline, confirming that explicitly modeling the relative priority among multiple positives leads to better sequence generation quality.

\subsection{Online A/B Test}
\label{sec:online}

TSGR is deployed as an additional channel alongside existing retrieval sources, with its outputs passed directly to the ranking stage, bypassing the conventional pre-ranking stage. The system is served on NVIDIA H20 GPUs with a latency budget of 65ms. Given a user query and context features, the model performs constrained beam search.
The first two SID levels are decoded under a valid-prefix constraint enforced by a pre-built prefix tree, which prunes invalid token sequences, while the third level uses standard beam search, yielding a set of complete SID sequences.
As each candidate is generated, the VRM computes a value score.
The candidates are then re-ranked and returned as the final output.

We deploy TSGR in a production A/B test on the Taobao search platform, serving 1\% of live traffic over 38 consecutive days. In the treatment group, TSGR replaces the conventional retrieval and pre-ranking stages with a single model, directly forwarding its top-1{,}500 ranked candidates to the ranking stage. Compared with the production baseline, TSGR achieves gains of \textbf{+0.43\%} in IPV, \textbf{+1.12\%} in Transaction Count, and \textbf{+1.64\%} in GMV. These improvements are statistically significant and, at the scale of Taobao Search, translate into substantial business value. Following the successful A/B test, TSGR has been fully deployed in production, serving hundreds of millions of users on the Taobao Search platform.

%% file: Tables/appendix_qpsid_hit.tex
\begin{table}
\centering
\caption{Hierarchical SID prediction accuracy of the top-100 output items under different label strategies.}
\label{tab:multipath_analysis}
\resizebox{\columnwidth}{!}{%
\begin{tabular}{lccc}
\toprule
Label Strategy & Cluster Hit Rate & Level-3 Hit Rate & Overall Hit Rate \\
\midrule
QP-SID (Row 1) & 78.71 & 72.83 & 57.32 \\
Multi-path (Row 3) & 77.20 & 71.90 & 55.50 \\
\bottomrule
\end{tabular}%
}
\end{table}

%% file: Tables/appendix_qpsid_ratio.tex
\begin{table}
\centering
\caption{Distribution of QPSID categories across training and evaluation sets, as well as model output.}
\label{tab:qpsid_distribution}
\small
\begin{tabular}{l cccc}
\toprule
\textbf{Category} & \textbf{Train Seq} & \textbf{Label} & \textbf{Eval Seq} & \textbf{Model Output} \\
\midrule
No Hit (Default) & 55.37\% & 23.20\% & 61.42\% & 32.05\% \\
Term Rank = 1    & 25.46\% & 54.54\% & 22.99\% & 44.91\% \\
Term Rank = 2    & 11.63\% & 15.00\% & 9.71\%  & 14.53\% \\
Term Rank = 3    & 7.54\%  & 7.26\%  & 5.88\%  & 8.51\%  \\
\bottomrule
\end{tabular}
\end{table}

%% file: Tables/sft_ablation.tex
\begin{table}
    \centering
    \caption{Ablation study on SFT training objectives.}
    \label{tab:sft_ablation}
    \begin{tabular}{lrrrrr}
        \toprule
        Method & HR@20 & HR@100 & HR@500 & \cellcolor{lightyellow}HR@1k & HR@5k \\
        \midrule
        Pointwise & 0.3851 & 0.5653 & 0.7188 & \cellcolor{lightyellow}0.7714 & 0.8577 \\
        Listwise  & \textbf{0.4042} & \textbf{0.5890} & \textbf{0.7405} & \cellcolor{lightyellow}\textbf{0.7922} & \textbf{0.8718} \\
        \bottomrule
    \end{tabular}
\end{table}

%% file: Sections/5_Conclusion.tex
\section{Conclusion}

We present TSGR, a unified generative retrieval framework that addresses the value-insensitivity of existing GR systems in industrial e-commerce search.
Our core insight is to embed value awareness into both item representation and model output: QP-SID encodes item efficiency directly into the semantic identifier, while a jointly optimized Value-aware Ranking Module enables a single model to serve as both retriever and pre-ranker, eliminating the objective misalignment of a separate pre-ranking stage.
To further align the model with business objectives, we adopt a training pipeline of Pre-SFT, weighted multi-positive SFT, and explore the role of RL.
Extensive offline experiments on Taobao search and online A/B tests demonstrate consistent and significant gains over strong baselines, and TSGR has been successfully deployed in production, serving real traffic at scale.

%% file: Sections/6_Appendix.tex
\section{Appendix}

\subsection{The Use of Large Language Models (LLMs)}
In this paper, we use the publicly available LLM Qwen as the backbone model of our experiments. As for writing, we use generative AI tools such as Grammarly solely to improve the clarity, coherence, and overall quality of the written text. All figures, preliminary drafts, and content are written independently by the authors without reliance on LLMs.

\subsection{Online Deployment}
\begin{figure}
    \centering
    \includegraphics[width=0.95\linewidth]{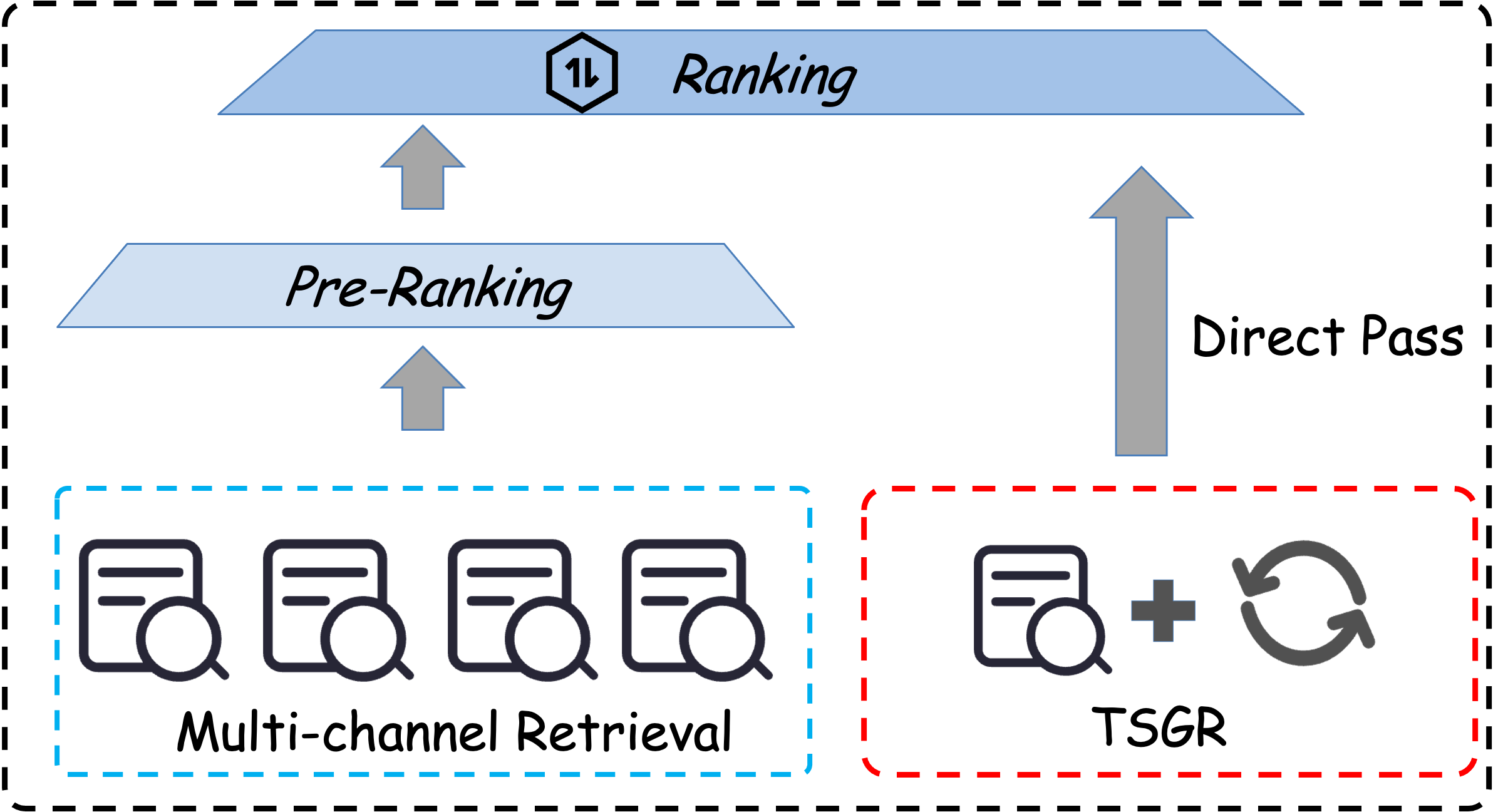}
    \caption{Online deployment architecture of TSGR. Items generated by TSGR bypass pre-ranking and enter the ranking stage directly.}
    \label{fig:architecture}
\end{figure}

As illustrated in Figure~\ref{fig:architecture}, TSGR is deployed as an independent retrieval channel alongside multi-channel retrieval in the Taobao Search pipeline, with the key distinction that its output bypasses pre-ranking and enters the ranking stage directly, given that the generated items are already ordered by value scores that reflect both relevance and commercial quality.

\subsection{Implementation Details}
\label{app:details}
The batch sizes for SFT and GRPO are 256 and 16 (2 per device ${\times}$ 8 gradient accumulation steps), respectively. SFT employs AdamW ($\beta_1{=}0.9$, $\beta_2{=}0.999$) with a cosine learning rate of $1{\times}10^{-4}$ and 10 warmup steps, whereas GRPO employs a linear learning rate of $4{\times}10^{-7}$ with 20 warmup steps over 2 epochs on 100K queries. Inference applies dynamic beam search with sizes $[400, 400, 10000]$ across the three SID levels, and $\lambda$ is set to 1. For GRPO, 32 candidates are sampled per query at a temperature of 1.0 under constrained decoding; the top-10 candidates by ranking CTR$\times$CVR score receive a reward of $+1$ and the bottom-10 receive a reward of $-1$, with 16 offline preference pairs appended per prompt. The KL coefficient is set to 0.5 and is scaled to $0.6{\times}$ when the KL and policy gradient directions conflict at the sample level. Entropy regularization ($0.005$) is applied to Levels~1--2 and disabled when the overlap exceeds $20\%$.

\subsection{QP-SID Case Study}
\label{app:QPSID}
\begin{figure*}
    \centering
    \includegraphics[width=0.98\linewidth]{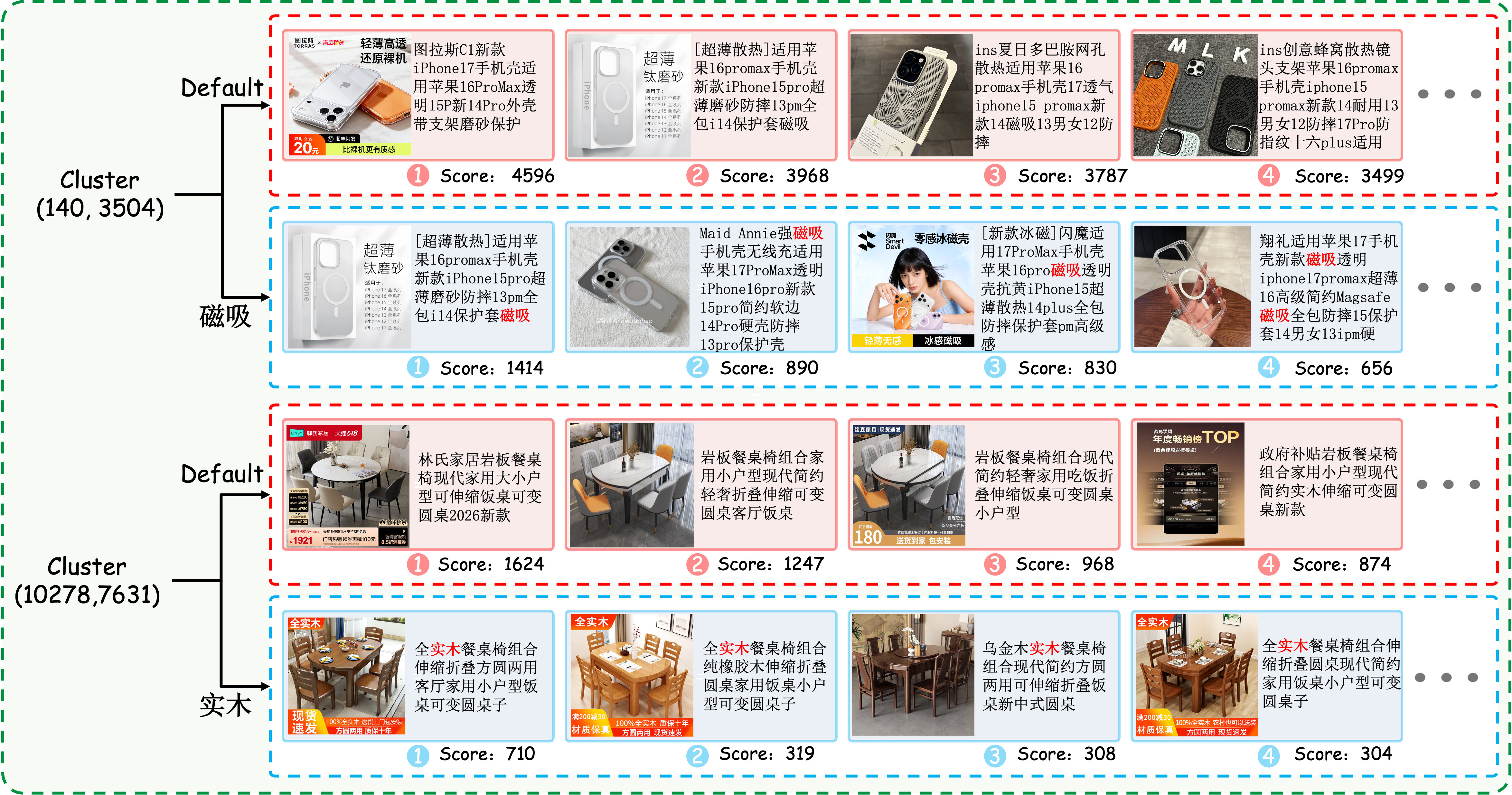}
    \caption{Case study of QP-SID on two clusters: phone cases (Cluster (140, 3504)) and dining tables (Cluster (10278, 7631)). For each cluster, we show the default value ordering (red) and the query-conditioned ordering under a specific term (blue), with matched terms highlighted in red. Scores beside each item denote the corresponding value scores.}
    \label{fig:case_1}
\end{figure*}

\begin{figure*}
    \centering
    \includegraphics[width=0.98\linewidth]{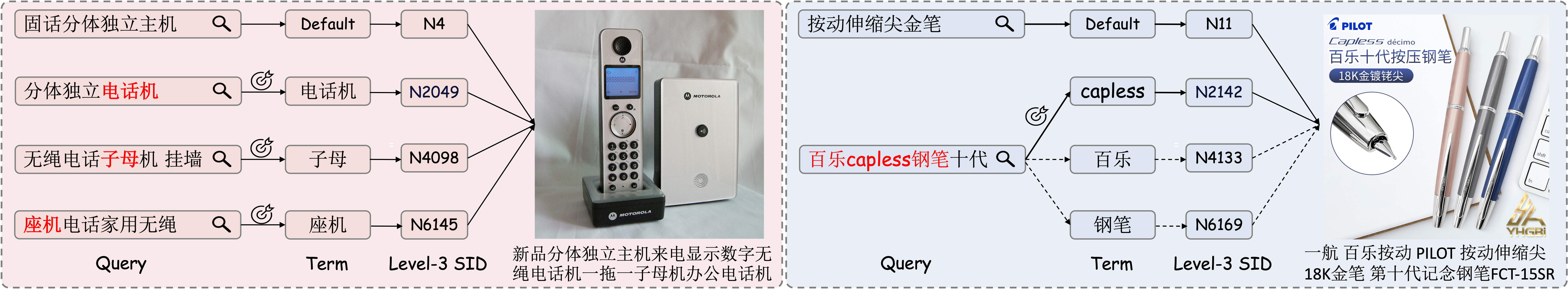}
    \caption{Case study of term selection in QP-SID during serving. \textbf{Left}: different queries each match a single term, which determines the corresponding Level-3 SID. \textbf{Right}: when a query matches multiple terms, QP-SID selects the top-ranked term (solid arrow) and discards the others (dashed arrows). Matched terms are highlighted in red.}
    \label{fig:case_2}
\end{figure*}

To intuitively illustrate how QP-SID organizes items, Figure~\ref{fig:case_1} presents a case study on two representative clusters: phone cases (Cluster (140, 3504)) and dining tables (Cluster (10278, 7631)).
For each cluster, we compare the \emph{default} value ordering with the \emph{query-conditioned} ordering under a representative term, namely ``\zh{磁吸}'' (magnetic) for phone cases and ``\zh{实木}'' (solid wood) for dining tables.
Items within each row are sorted in descending order of their value scores.
Note that the default and query-conditioned scores are computed under different criteria and are therefore not directly comparable; only the relative ordering within each row is meaningful.

The comparison reveals two complementary behaviors of QP-SID.
The default ordering reflects the overall commercial value of items within a cluster, ranking globally popular and high-value items first.
In contrast, the query-conditioned ordering re-prioritizes items according to query intent: under the term ``\zh{磁吸}'' or ``\zh{实木}'', items whose titles explicitly match the term (highlighted in red) are promoted to the top positions, even if they rank lower in the default ordering.
This demonstrates that QP-SID maintains multiple query-aware value orderings for the same cluster, enabling the model to select the ordering that best matches the dominant intent of a given query and thus place intent-aligned high-value items in more favorable token positions.

Figure~\ref{fig:case_2} further illustrates how QP-SID selects among these orderings during serving, where each ordering is indexed by a term derived from term-item click statistics (Section~\ref{sec:parallel_sid}).
We consider two representative cases.
In the single-match case (Figure~\ref{fig:case_2}, left), each query matches exactly one term, which directly determines the Level-3 SID assigned to the target item.
For example, the queries ``\zh{分体独立电话机}'' and ``\zh{座机电话家用无绳}'' match the terms ``\zh{电话机}'' and ``\zh{座机}'', respectively, showing that semantically distinct queries pointing to the same item can be routed through different terms, each reflecting the specific intent of the query.
In the multi-match case (Figure~\ref{fig:case_2}, right), a query may match several terms simultaneously, and QP-SID selects the top-ranked term to govern the value ordering while discarding the rest.
For instance, the query ``\zh{百乐capless钢笔十代}'' matches three terms (``capless'', ``\zh{百乐}'', and ``\zh{钢笔}''), and the top-ranked term ``capless'' (solid arrow) is selected while the others are discarded (dashed arrows).
Taken together, these cases show that QP-SID flexibly adapts its value ordering to query intent, ensuring that the most intent-relevant ordering guides SID generation during serving.

\subsection{Prompt Templates for the Pipeline}
\input{Tables/appendix_prompt}
\label{app:prompts}
Table~\ref{tab:prompt-templates} presents the templates for the five Pre-SFT task types, and Table~\ref{tab:sft-prompt} presents the unified template shared by the SFT and RL stages.
Each template specifies the natural-language instruction together with placeholders, denoted by \texttt{\$\{$\cdot$\}}, that are filled with the corresponding item attributes (\eg, title, semantic ID) or user attributes (\eg, gender, age, purchasing power, historical interaction sequence) at training time.
All prompts are originally written in Chinese, consistent with the Taobao Search scenario, and are presented here in their original form to faithfully reflect the actual model inputs.

\subsection{Details of RL}
\label{app:rl}
\input{Tables/rl}
\paragraph{Reward Design.}
The reward steers generation toward candidates that are both relevant and high in predicted business value, defined as the product of CTR and CVR from the ranking model:
\begin{equation}
    r(s) = \mathrm{CTR}(q,\, i_s) \times \mathrm{CVR}(q,\, i_s).
\end{equation}

\paragraph{Offline DPO Augmentation.}
Standard GRPO relies on on-policy rollouts to form group-relative rewards, but these rollouts are typically concentrated on head candidates. As a result, reward normalization systematically suppresses lower-reward samples, even though some of them may still fall within the top-$1000$ region of interest. While this bias may benefit head-heavy metrics, it is suboptimal for our target retrieval objective. To mitigate this issue, we augment each prompt with offline preference pairs constructed from ranking CTR$\times$CVR scores: chosen samples are drawn from the ranking top-$B$ candidates and ground-truth purchase labels, while rejected samples are sampled from candidates around rank $1000$ to reduce the risk of false negatives. Furthermore, after mixing the on-policy rollouts with the offline preference samples, we retain only the top-$k$ and bottom-$k$ samples in the merged set by reward ($k=10$), and mask out the rewards of all remaining samples. This design concentrates optimization on high-confidence preference signals while preventing GRPO's group-relative normalization from mistakenly suppressing ambiguous candidates that may still belong to the top-$1000$ set.

\paragraph{Dynamic Hierarchical Entropy Regularization.}
To encourage exploration of diverse SID paths, we apply entropy regularization with a dynamic, layer-wise strategy.
For each prompt, we measure the match rate $m$ between the rollout SIDs and the ground-truth SIDs at Levels~1--2, and gate the regularization with a binary coefficient:
\begin{equation}
    \beta(m) =
    \begin{cases}
        \beta_0 & \text{if } m < \tau_{\text{low}}, \\
        0       & \text{if } m > \tau_{\text{high}},
    \end{cases}
\end{equation}
so that exploration is restricted to prompts that have not yet converged.
We apply it only to Levels~1--2, since entropy at Level~3 would perturb the fine-grained item distribution calibrated by SFT.
The resulting term is added to the GRPO objective:
\begin{equation}
    \mathcal{L}_{\text{RL}} = \mathcal{L}_{\text{GRPO}} - \beta(m) \sum_{l \in \{1, 2\}} \mathcal{H}(\pi_\theta^{(l)}),
\end{equation}
where $\mathcal{H}(\pi_\theta^{(l)})$ denotes the entropy of the policy distribution at Level $l$.

Table~\ref{tab:ablation_unified} shows that each RL component contributes incrementally. Basic RL alone yields only marginal gains ($+0.02\%$). DPO Augmentation brings a more notable improvement ($+0.39\%$), indicating that preference-based supervision helps the model better distinguish between positive and negative candidates. Dynamic Entropy Regularization further improves performance ($+0.11\%$), confirming that maintaining output diversity during RL training is beneficial. The modest cumulative gain of $+0.52\%$ explains why RL has not been adopted in actual deployment.
\subsection{TencentGR Dataset}
\label{app:tencent_gr}
\begin{table}[t]
\centering
\caption{Ablation experiment on the TencentGR-10M dataset.}
\label{tab:tencent}
\begin{tabular}{l l c}
\toprule
\textbf{Dataset} & \textbf{Method} & \textbf{HR@500} \\
\midrule
\multirow{2}{*}{TencentGR-10M} & Base & 0.0786 \\
& \textbf{Value-aware Ranking Module} & \textbf{0.0794} \\
\bottomrule
\end{tabular}
\end{table}
\input{Tables/all_rank}
The TencentGR dataset~\cite{pan2026tencent} poses several practical challenges that require dedicated preprocessing before it can be used to evaluate our framework.
Specifically, the dataset provides no query field, no pre-built semantic IDs, fully anonymized user and item attributes, and no public test set with ground-truth labels.
We describe below how each challenge is addressed.

\noindent\textbf{SID Construction.}
Since the dataset does not provide semantic IDs, we construct them directly from the 1024-dimensional multimodal item embeddings.
We apply hierarchical clustering to produce a two-level SID prefix, and determine the third level by ranking items within each cluster in descending order of their transaction, click, and impression counts.
Each item is thus mapped to a three-level SID.

\noindent\textbf{Item-Side Information.}
As raw item content is not publicly available, item-side information is assembled from three sources: (1) the constructed SID; (2) anonymized item features concatenated with per-item impression, click, and transaction counts aggregated from the behavior sequences (missing values are filled with $-1$); and (3) the $\ell_2$-normalized 1024-dimensional multimodal embedding.

\noindent\textbf{Behavior Sequence Preprocessing.}
The raw behavior sequences contain approximately 10.1M records, many of which consist entirely of impressions with no clicks or transactions and are therefore uninformative for training or evaluation.
After filtering out these impression-only records, approximately 7.71M records are retained.
Each remaining record is then processed for multi-label training: the most recent impression, click, and transaction are each selected as a label; to prevent label leakage, only behavior that occurs before the earliest label timestamp is used as input; and the input is split into three separate sub-sequences corresponding to impressions, clicks, and transactions, respectively.

\noindent\textbf{Prompt Construction.}
As no interpretable user attributes are available, all user features are included in the prompt in the form \texttt{Feature $N$: value} without semantic distinction.
The prompt instructs the model to recommend the next item given the user features and the three historical sub-sequences encoded as SID strings.
The output consists of up to three SID strings corresponding to the transaction, click, and impression labels in order of priority, with lower-priority labels used as fallbacks when higher-priority ones are absent.
Each training sample additionally carries the multimodal embedding and side information of the label items, a label priority indicator, and group identifiers used for multi-label alignment.

\noindent\textbf{Train/Test Split.}
As no public test set is available, we construct a reproducible split by sorting all records by MD5(user\_id) and designating the first 20,000 records as the test set, with the remainder used for training.
This procedure ensures that users in the test set do not appear in the training set.

\subsection{Other Experiments}
\label{app:exp}
% Table~\ref{tab:qpsid_distribution} presents the distribution of QP-SID categories across different data splits. The majority of items in both training and evaluation sequences fall into the No Hit (Default) category, indicating that most historically clicked items in user behavior sequences are not closely related to the current query. In contrast, training labels show a markedly different distribution, where up to 76.8\% of clicked items are query-relevant (Term Rank = 1, 2, or 3), which is expected since users click with the current query intent in mind. The model output distribution further suggests that the model has successfully learned to retrieve candidates from multiple ranking channels rather than relying solely on the Default channel.

% To further diagnose the performance degradation under the Multi-path label strategy (Row 3 of Figure~\ref{fig:sid}), we analyze the hierarchical prediction accuracy across all three SID levels (Table~\ref{tab:multipath_analysis}). Introducing auxiliary query-matched labels increases the learning difficulty of the third level, amplifying its loss and gradient and propagating the interference to the first two levels. Moreover, 99.8\% of the model outputs collapse to the Default path, indicating that the model fails to leverage the diversity of QP-SID orderings and degenerates into a single-path retrieval system. These results confirm that auxiliary labels undermine the learning of all three levels, which explains the consistent performance drop observed across retrieval metrics.

We provide a comprehensive ablation study of the value-aware ranking module in Table~\ref{tab:ranking_ablation}. We evaluate different ranking strategies, including PV, CTR, CVR and their combinations, across multiple model variants. Results show that the Attention-based model with $PV \times CTR$ ranking achieves the best overall performance.

%% file: Tables/appendix_prompt.tex
% ⭐ 关键：让 X 列垂直居中
\renewcommand\tabularxcolumn[1]{m{#1}}

% ⭐ 关键：让 X 列垂直居中
\renewcommand\tabularxcolumn[1]{m{#1}}

\begin{table*}[t]
\centering
\caption{Prompt templates for the five Pre-SFT task types.}
\label{tab:prompt-templates}
\renewcommand{\arraystretch}{1.5}
\setlength{\extrarowheight}{2pt}
\begin{tabularx}{\textwidth}{C L}
\toprule
\rowcolor{headblue}
{\centering\color{white}\bfseries Task Type\par} &
{\centering\color{white}\bfseries Task Definition\par} \\
\midrule

\textbf{Encoding} &
\zh{商品的标题为：}\texttt{\$\{title\}}\zh{，请预测它的语义id} \\
\midrule

\rowcolor{rowgray}
\textbf{Personalization} &
\zh{你是一个电商搜索引擎，你的任务是根据用户的个人信息、当前的搜索词、用户历史交互的商品序列，为用户检索下一个商品。用户的性别为：}\texttt{\$\{sex\}}\zh{，年龄为：}\texttt{\$\{age\}}\zh{，购买力级别为：}\texttt{\$\{level\}}\zh{。用户历史交互的商品序列为：}\texttt{\$\{sids\}}\zh{。你检索的商品标题是} \\
\midrule

\textbf{Retrieval} &
\zh{你是一个电商搜索引擎，你的任务是根据当前的搜索词，为用户检索商品。搜索词为：}\texttt{\$\{query\}}\zh{。你检索的商品是} \\
\midrule

\rowcolor{rowgray}
\textbf{Collaboration} &
\zh{某商品的标题为：}\texttt{\$\{trigger\_title\}}\zh{，它的语义id为：}\texttt{\$\{trigger\_sid\}}\zh{，购买过它的用户通常还购买过相似商品，标题为：}\texttt{\$\{target\_title\}}\zh{，请预测这一相似商品的语义id} \\
\midrule

\textbf{Recommendation} &
\zh{你是一个电商推荐引擎，你的任务是根据用户的个人信息、用户历史交互的商品序列，为用户推荐下一个商品。用户的性别为：}\texttt{\$\{sex\}}\zh{，年龄为：}\texttt{\$\{age\}}\zh{，购买力级别为：}\texttt{\$\{level\}}\zh{。用户历史交互的商品序列为：}\texttt{\$\{sids\}}\zh{。你推荐的商品是} \\

\bottomrule
\end{tabularx}
\end{table*}

% SFT prompt
\begin{table*}[t]
\centering
\caption{SFT and RL prompt template for personalized retrieval.}
\label{tab:sft-prompt}
\renewcommand{\arraystretch}{1.5}
\setlength{\extrarowheight}{2pt}
\begin{tabularx}{\textwidth}{C L}
\toprule
\rowcolor{headblue}
\multicolumn{1}{>{\centering\arraybackslash}m{2.6cm}}{\textcolor{white}{\bfseries Role}} &
\multicolumn{1}{>{\centering\arraybackslash}m{\dimexpr\linewidth-2.6cm-4\tabcolsep\relax}}{\textcolor{white}{\bfseries Content}} \\
\midrule

\textbf{System} &
\zh{请根据用户信息、搜索词、用户历史交互商品序列，为用户检索下一个商品。} \\
\midrule

\rowcolor{rowgray}
\textbf{User} &
\zh{用户性别为：}\texttt{\$\{gender\}}\zh{，年龄为：}\texttt{\$\{age\}}\zh{，购买力级别为：}\texttt{\$\{level\}}\zh{。搜索词为：}\texttt{\$\{query\}}\zh{。历史交互商品序列为：}\texttt{\$\{opt\_seq\}}\zh{。你检索的商品是} \\
\bottomrule
\end{tabularx}
\end{table*}

%% file: Tables/rl.tex
% 导言区添加：
% \usepackage[table]{xcolor}
% \definecolor{lightyellow}{RGB}{255,255,204}

\begin{table}
  \centering
  \footnotesize
  \setlength{\tabcolsep}{4pt}
  \caption{Ablation study of RL strategies. Each row cumulatively adds one component upon the previous configuration.}
  \label{tab:ablation_unified}
  \begin{tabular}{lccccc}
    \toprule
    Method & HR@20 & HR@100 & HR@500 & \cellcolor{lightyellow}HR@1k & HR@5k \\
    \midrule
    Baseline
      & 0.4965 & 0.6580 & 0.7746 
      & \cellcolor{lightyellow}0.8111 & 0.8747 \\
    \quad + Basic RL
      & 0.4985 & 0.6584 & 0.7761 
      & \cellcolor{lightyellow}0.8113 & 0.8737 \\
    \quad + DPO Augmentation
      & 0.4975 & 0.6658 & 0.7803 
      & \cellcolor{lightyellow}0.8152 & 0.8772 \\
    \quad + Dynamic Entropy Reg
      & \textbf{0.4989} & \textbf{0.6674} & \textbf{0.7808}
      & \cellcolor{lightyellow}\textbf{0.8163} & \textbf{0.8788} \\
    \bottomrule
  \end{tabular}
\end{table}

%% file: Tables/all_rank.tex
\begin{table*}[t]
\centering
\caption{Ablation study of the Value-aware Ranking Module.}
\label{tab:ranking_ablation}
\resizebox{0.85\textwidth}{!}{
\begin{tabular}{l l ccccc}
\toprule
\textbf{Model Design} & \textbf{Ranking Strategy}
& \textbf{HR@20} & \textbf{HR@100} & \textbf{HR@500} & \cellcolor{lightyellow}\textbf{HR@1000} & \textbf{HR@5000} \\
\midrule
\multirow{6}{*}{Attention}
& Base             & \textbf{0.4754} & 0.6431 & 0.7807 & \cellcolor{lightyellow}0.8252 & 0.8953 \\
& PV               & 0.3387 & 0.5803 & 0.7889 & \cellcolor{lightyellow}0.8482 & 0.8953 \\
& CTR              & 0.3744 & 0.5401 & 0.7088 & \cellcolor{lightyellow}0.7824 & 0.8953 \\
& CVR              & 0.2149 & 0.3097 & 0.4717 & \cellcolor{lightyellow}0.5802 & 0.8953 \\
& PV$\times$CTR          & 0.4586 & \textbf{0.6677} & \textbf{0.8250} & \cellcolor{lightyellow}\textbf{0.8651} & 0.8953 \\
& PV$\times$CTR$\times$CVR     & 0.4443 & 0.6626 & 0.8239 & \cellcolor{lightyellow}0.8650 & 0.8953 \\
\midrule
\multirow{6}{*}{Mean Pooling}
& Base             & 0.4774 & 0.6492 & 0.7822 & \cellcolor{lightyellow}0.8269 & 0.8985 \\
& PV               & 0.2726 & 0.5273 & 0.7599 & \cellcolor{lightyellow}0.8303 & 0.8985 \\
& CTR              & 0.3148 & 0.4995 & 0.7020 & \cellcolor{lightyellow}0.7865 & 0.8985 \\
& CVR              & 0.1863 & 0.3002 & 0.4825 & \cellcolor{lightyellow}0.6008 & 0.8985 \\
& PV$\times$CTR          & 0.3920 & 0.6161 & 0.8003 & \cellcolor{lightyellow}0.8488 & 0.8985 \\
& PV$\times$CTR$\times$CVR     & 0.3849 & 0.6146 & 0.7999 & \cellcolor{lightyellow}0.8503 & \textbf{0.8985} \\
\midrule
\multirow{6}{*}{EOS Token}
& Base             & 0.4763 & 0.6493 & 0.7840 & \cellcolor{lightyellow}0.8272 & 0.8948 \\
& PV               & 0.3587 & 0.5975 & 0.7942 & \cellcolor{lightyellow}0.8483 & 0.8948 \\
& CTR              & 0.3455 & 0.5278 & 0.7227 & \cellcolor{lightyellow}0.7977 & 0.8948 \\
& CVR              & 0.2418 & 0.3662 & 0.5448 & \cellcolor{lightyellow}0.6514 & 0.8948 \\
& PV$\times$CTR          & 0.3862 & 0.6058 & 0.8016 & \cellcolor{lightyellow}0.8539 & 0.8948 \\
& PV$\times$CTR$\times$CVR     & 0.3727 & 0.5916 & 0.7953 & \cellcolor{lightyellow}0.8512 & 0.8948 \\
\midrule
\multirow{6}{*}{No Embedding}
& Base             & 0.4775 & 0.6432 & 0.7790 & \cellcolor{lightyellow}0.8233 & 0.8954 \\
& PV               & 0.2236 & 0.4016 & 0.6336 & \cellcolor{lightyellow}0.7390 & 0.8954 \\
& CTR              & 0.2123 & 0.3342 & 0.5092 & \cellcolor{lightyellow}0.6053 & 0.8954 \\
& CVR              & 0.1548 & 0.2355 & 0.3656 & \cellcolor{lightyellow}0.4613 & 0.8954 \\
& PV$\times$CTR          & 0.2600 & 0.4356 & 0.6448 & \cellcolor{lightyellow}0.7487 & 0.8954 \\
& PV$\times$CTR$\times$CVR     & 0.2621 & 0.4384 & 0.6483 & \cellcolor{lightyellow}0.7502 & 0.8954 \\
\midrule
\multirow{6}{*}{No Side-Info}
& Base             & 0.4763 & 0.6469 & 0.7803 & \cellcolor{lightyellow}0.8269 & 0.8963 \\
& PV               & 0.3868 & 0.6097 & 0.7922 & \cellcolor{lightyellow}0.8429 & 0.8963 \\
& CTR              & 0.3580 & 0.5382 & 0.7194 & \cellcolor{lightyellow}0.7913 & 0.8963 \\
& CVR              & 0.1645 & 0.2404 & 0.4096 & \cellcolor{lightyellow}0.5432 & 0.8963 \\
& PV$\times$CTR          & 0.4116 & 0.6412 & 0.8139 & \cellcolor{lightyellow}0.8580 & 0.8963 \\
& PV$\times$CTR$\times$CVR     & 0.4013 & 0.6341 & 0.8125 & \cellcolor{lightyellow}0.8570 & 0.8963 \\
\bottomrule
\end{tabular}
}
\end{table*}